# Correlation of interfacial bonding mechanism and equilibrium conductance of molecular junctions

Zhanyu NING (宁展宇)[2], Jingsi QIAO (乔婧思)[1], Wei JI (季威)[1,2,*] and Hong GUO (郭鸿)[2,†]

[1]Department of Physics and Beijing Key Laboratory of Optoelectronic Functional Materials & Micro-Nano Devices, Renmin University of China, Beijing 100872, China

[2]Centre for the Physics of Materials and Department of Physics, McGill University, Montreal, QC, Canada H3A 2T8

Email: [*]wji@ruc.edu.cn (W.J.); [†]guo@physics.mcgill.ca (H.G.)

**Abstract**

We report theoretical investigations on the role of interfacial bonding mechanism and its resulting structures to quantum transport in molecular wires. Two bonding mechanisms for the Au-S bond in an Au(111)/1,4-benzenedithiol(BDT)/Au(111) junction were identified by *ab initio* calculation, confirmed by a recent experiment, which, we showed, critically control charge conduction. It was found, for Au/ BDT/Au junctions, the hydrogen atom, bound by a dative bond to the Sulfur, is energetically non-dissociative after the interface formation. The calculated conductance and junction breakdown forces of H-non-dissociative Au/BDT/Au devices are consistent with the experimental values, while the H-dissociated devices, with the interface governed by typical covalent bonding, give conductance more than an order of magnitude larger. By examining the scattering states that traverse the junctions, we have revealed that mechanical and electric properties of a junction have strong correlation with the bonding configuration. This work clearly demonstrates that the interfacial details, rather than previously believed many-body effects, is of vital importance for correctly predicting equilibrium conductance of molecular junctions; and manifests that the interfacial contact must be carefully understood for investigating quantum transport properties of molecular nanoelectronics.

## 1. INTRODUCTION

In a very recent paper, Song *et al.* [1] reported experimental fabrications of three-terminal single molecule field effect transistors. The successful fabrication and characterization of such a molecular device can be considered an important milestone of nanoelectronics. In the experiment, a gold wire was broken by electro-migration to produce a nano-meter gap in the wire. A molecule such as 1,4-benzenedithiol (BDT) may bridge the gap and form an Au/BDT/Au single molecule transport junction. A range of equilibrium conductance of 0.011~ 0.015 $G_0$ ($G_0 = 2e^2/h$) was reported giving an average value of 0.0132 ± 0.0021 $G_0$. The Au/BDT/Au device was subjected to extensive studies in the past, for instance, a mean conductance of 0.011 $G_0$ was reported by statistically measuring several thousand Au/BDT/Au junctions formed by BDT bridging the gap between a scanning tunneling microscope (STM) tip and a Au surface [2]. Their results have also been reproduced by other experiments [3]. The consistency of measured transport properties on devices fabricated by totally different methods suggests a degree of structural-function control at the single molecule level.

Despite these experimental achievements, the atomic structures of the fabricated devices, especially at the interface, were still unclear. In particular, the most important structural information, namely the metal-molecule contacts, is at best ambiguous for essentially all single molecule transport junctions investigated in the literature. In general, a most important science issue concerning nano-systems is the relationship between structure and function. The experimental convergence of transport data for the Au/BDT/Au device provides a timely opportunity to shed considerable light on the structure-function issue of molecular nanoelectronics. It is the purpose of this paper to report our first principles theoretical investigation on how the contacts can critically affect quantum transport properties of molecular junctions. In a typical theoretical analysis of molecular devices, one assumes an initial contact structure between the molecule and the metal electrodes, which is guided, usually, by intuition or by what people believe to be, and then, in few studies, relaxes the structure. However, experimentally when a molecule is brought to contact the metal leads, a contact formation occurs where chemical reactions may give rise to dissociation or formation of atomic groups from the original molecule. Such a process is likely lost when an initial atomic configuration is *assumed* without carefully considering it from *ab initio* point of view. As a result, the bonding mechanism governed interfacial geometry after the formation of contact has not been subjected to systematic investigations so far and, as we show below, it is a crucial effect that controls the interface transparency to charge flow.

We use the Au(111)/1,4-benzenedithiol(BDT) / Au(111) junction as a prototypical system for our investigation. As discussed above, the experimental conductance value is 0.011-0.015 $G_0$ [1,2] for this system. On the theoretical side [4-15], conductance obtained from quantitative analysis have not been able to reach a clear consensus. Density functional theory (DFT) based first principles methods within local density approximation (LDA) or generalized gradient approximation (GGA) have mostly produced conductance values considerably larger, by more than one or even two orders of magnitude, than the converged experimental value of 0.011-0.015 $G_0$ [1,2]. Many-body electronic correlation effects was suggested can alter the conductance of molecular junctions by an order of magnitude [16] and was illustrated, with an example of Au/1,4-benzenediamine(BDA)/Au junctions, crucial in properly predicting equilibrium conductance of molecular junctions. However, the consideration of many-body effects for the Au/BDT/Au junction has not been examined for some reason.

In this paper, we focus on a different issue, namely the role of interfacial geometry at the contact in transport properties of molecular junctions, by massive first principles calculations, we found another bonding configuration which is more energetically favored than the previously proposed one [17] for the Au/BDT/Au junction. In this configuration, a BDT molecule prefers to attach to ad-atoms when bonded to Au(111) and, more importantly, the hydrogen atoms on the thiol groups of a BDT do not dissociate away after the formation of the device contacts. The breakdown force [18] of this energetically more stable configuration, denote H-non-dissociated (HND) model, compares better to the experimental data than that of the previous H-dissociated (HD) model. The calculated conductance of the junctions with HND model is within a factor of two to five in comparison to experimentally measured values. On the other hand, all the HD configurations produce conductance at least one order of magnitude greater. This rather large discrepancy of conductance is found a result of distinctly different bonding mechanism, namely dative bonding filled by a lone pair of S and covalent bonding filled by electrons from Au, as suggested by analyzing the scattering states that traversing the metal-molecule contacts.

## 2. METHODS

### 2.1 Total energy calculation

Density functional theory (DFT) total energy calculations were carried out using general gradient approximation for the exchange-correlation potentials (Perdew-Burke-Ernzerhof, PBE) [19], the projector augmented wave method [20] (PAW), and a plane wave basis set up to 400 eV as implemented in the Vienna *ab-initio* Simulation Package VASP [21]. A c(4×3) supercell consists of six layers of Au atoms separated by a vacuum layer of 15 Å was adopted to model the Au/BDT interface. All atoms except the three bottom Au layers were fully relaxed with a force criterion of 0.02 eV/Å applied for every ion. A plane wave kinetic energy cutoff of 400 eV and a *k*-mesh of 4×4×1 were further checked by a cutoff of 500 eV and a mesh of 8×8×1 to ensure the convergence to 1 meV/atom.

### 2.2 Modeling of junction in structural evolution

Junctions were modeled using the ``Dual-Slab Model'', proposed by this work, in terms of structural relaxation. With this model, the fully relaxed structure of a molecular junction can be easily accessed with smaller computational effort. In this model, two slabs, consisting four Au layers for each slab and separated by two vacuum regions, were employed to represent two semi- infinite leads, see Figure 1(a). Here, there is an assumption taken from the field of surface modeling that a minimal thickness of four layers of metal atoms is able to capture the main feature of a semi-infinite surface. In each slab of four-layer Au atoms, we used the two outer layers, which were kept fixed during structural relaxations, to simulate the bulk atoms and the two inner layers for surface atoms. Such a setup ensures that the volume of the supercell is a constant during junction elongation, which avoids the Pulay force error with non-absolutely-converged basis sets. While for transport calculation, the central scattering region of the two-probe model could be directly derived from this dual-slab model by removing the right vacuum layer shown in Fig. 1(a).

Two adatoms are available on two face-to-face surfaces of the two slabs, while BDT molecules were put in between the two adatoms. Junction gap width $L$ is defined as the gap between the two face-to-face surfaces at their initial positions (in Au bulk positions). The difference of $z$ coordination of the adatom and the third layer (the top-most constrained layer during structural relaxations) defines $L_L$ ($L_R$) for the left (right) lead. Variable $D$ has a similar definition as that of $L_L$ or $L_R$, but solely for a lead, see Fig. 1(b).

At each $L$, the position of every atom in the dual-slab, except the four ``bulk'' layers (two outer layers in each slab), was fully relaxed using the PAW method. The cross-section and other details, e.g. the force criterion, in dual-slab calculations are the same as that for Au/BDT interfaces. To simulate the stretching process of the junction, one Au slab was moved step by step with a step length of 0.1 Å. A total number of 22 configurations, either in HND or HD model, have been considered from $L$ = 12.7 Å to 15.8 Å (16.7 Å for the HD model), which covers the length range from compressed to break-down junctions. The size of the supercell was kept fixed throughout the stretching process.

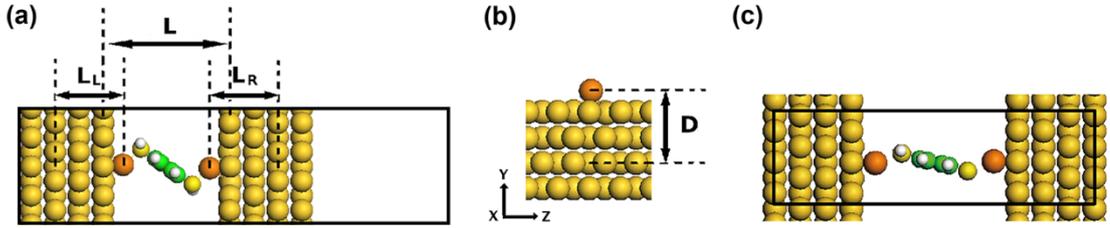

Fig. 1 Schematic diagrams for the ``dual-slab'' model with the definitions of $L$, $L_L$ and $L_R$ (a), the surface model with an adatom on a bare Au(111) surface showing the definition of $D$ (b), and the central scattering region of the ``two-probe'' model.

**2.3 Breakdown force calculation**

Breakdown force is defined as the difference of the measured forces (tension) on a lead of the junction just before and after the breakdown. The forces were experimentally recorded from atomic force microscope (AFM) tips in the AFM-Molecule-Surface setup which was adopted to measure the breakdown force of Au-S linkers [18]. Tensions before and after the breakdown were calculated using two methods, respectively. For the tension of a lead before the breakdown $T_{before}$, it equals to that of the whole junction, i.e. the tension of a dual-slab model. It is thus derived from the total energies of the junctions with various lengths, read as $T_{before}$ = d$E$/d$L$ ≈ $\Delta E/\Delta L$ at the junction gap right before the breakdown. Note that, the zero tension is defined by the configuration that gives the lowest total energy of the junction, in which the lead has already been stretched. Following the definition, it should be negative that the tension in a bare lead, as denoted $T_{after}$. Such force was calculated according to the $z$ position of an adatom solely on a bare Au surface ($D$). At the equilibrium $L_0$, the optimized distance $D$ of the adatom $D_{equilibrium}$ is

0.088 Å larger than the $D_{bare}$ on a bare lead. Tension $T_{after}$ is thus roughly equal to the Hellmann-Feynman force of the adatom directly calculated from the total energy of a lead, in which $D$ is increased to match $D_{equilibrium}$. The breakdown force is, by definition, read as $F_{breakdwon} = T_{before} - T_{after}$.

**2.4 Electron transport calculation**

Our transport analysis is based on carrying out DFT analysis within the Keldysh non-equilibrium Green's function (NEGF) formalism. The basic idea of NEGF-DFT is to self-consistently calculate the Hamiltonian of the device by DFT and determine the non-equilibrium quantum statistical properties of the device operation by NEGF. For more details we refer interested readers to the original literature. In the NEGF-DFT self-consistent calculation of the density matrix and Hamiltonian, we use double-$\zeta$ plus polarization (DZP) linear combination of atomic orbital (LCAO) basis sets for all the atoms, GGA-PBE for the exchange-correlation potential, and define atomic core potentials using standard norm conserving pseudopotentials. Each basis set was fully optimized using Nanobase, a package disturbed with Nanodcal, to ensure it is consistent with the corresponding PAW calculation, with the maximum energy difference of less than 0.1 eV in the comparison of bandstructures.

A two-probe model of Au/BDT/Au device is comprised of three parts [22,23], the scattering region plus the left/right leads. The scattering region, built directly from the relaxed structure in the dual-slab model, includes a BDT molecule bonded to four layers of Au(111) atoms via an ad-atom on each side (atoms inside the black rectangular box in Fig. 1(c)). The leads are bulk-like Au(111) layers extending on each side of the scattering region to $z=\pm\infty$ where $z$ is the direction of current flow. Periodic boundary condition is applied in $x$ and $y$ directions.

## 3. RESULTS AND DISCUSSION

### 3.1 Contact at interface

Despite the importance of bonding structure of metal-molecule interfaces to transport [24,25], detailed bonding geometry of Au/BDT/Au junctions is yet to be well established. In different systems such as the popular Au/alkanethiol/Au junction, the conductance is dominated by electron tunneling through localized $\sigma$ bonds of alkanes [26], namely the huge conductance (typically tens of mega-Ohms) of the junction is dominated by the length of the alkane molecule and not by the Au/alkanthiol interface. For Au/BDT/Au junctions, the benzene ring in BDT consists of a non-local $\pi$ bond system which is rather transparent to electron conduction hence the conductance should be much more sensitive to the bonding geometry at the Au/BDT interface. A very careful calculation on this interface is thus necessary. As suggested by a recent STM experiment [27], thiol-groups prefer to attach Au(111) surface through Au ad-atoms. We have therefore calculated BDT absorption on Au(111) with and without Au ad-atoms. Our total energy calculation suggests that BDT tends to attach the Au surface via adatoms, at least 0.4 eV per molecule more stable than the corresponding interface without adatom. As a result we shall focus on a series of representative atomic configurations with BDT molecules absorbed on Au(111) surface via Au ad-atoms.

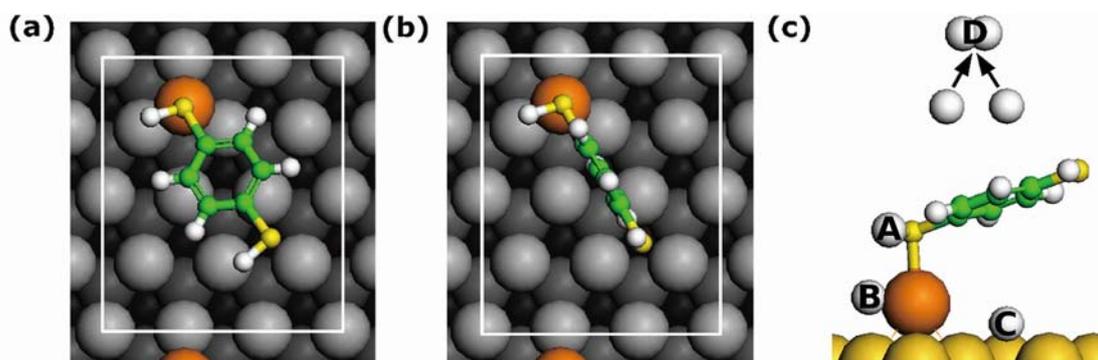

Fig. 2 Top views of examples for (a) parallel and (b) perpendicular configurations. (c) Four positions of H have been considered for each configuration at several orientations: (A) is for non-dissociated H; (B,C,D) are for dissociated H which attaches to the (B) ad-atom, (C) the surface, or (D) escapes into vacuum forming an $H_2$ molecule.

Figure 2(a) and (b) plot two initial structures of BDT attached to Au(111) via an ad-atom prepared for the structural relaxations, where the molecule is parallel (perpendicular) to the surface. According to the $C_2$ and $C_3$ rotational symmetries of the molecule and Au(111), another two sets of these structures with a rotation angle of -30° and 30° were also considered. These three sets should likely cover most of the initial configurations. For the interfacial contact, we consider one case for the non-dissociative contact, as suggested by the recent experimental [28] and theoretical [29] studies, and three cases for the dissociative contact. They are, i) the H remains attached to the S atom (A in Figure 2(c)), ii) the dissociated H attaches to the ad-atom (B), iii) to the surface (C), and escapes to vacuum to form an $H_2$ molecule (D), as shown in Figure 2(c).

Table I Difference of total energies (units eV) compared with the most stable structure (the parallel 0˚) for typical configurations of Au/BDT interface. For all situations, the H-non-dissociative structures have lower energies.

|            | Parallel |      |      | Perpendicular |      |      |
|------------|------|------|------|------|------|------|
|            | 0°   | 30°  | 60°  | 0°   | 30°  | 60°  |
| HND        | 0.00 | 0.09 | 0.08 | 0.06 | 0.06 | 0.06 |
| HD-adatom  | 0.28 | 0.25 | 0.61 | 0.53 | 0.54 | 0.42 |
| HD-surface | 0.41 | 0.40 | 0.42 | 0.40 | 0.41 | 0.41 |
| HD-vacuum  | 0.24 | 0.25 | 0.26 | 0.25 | 0.25 | 0.24 |

Table I summarizes the total energy of these interfaces. It is striking to find that configurations with H non-dissociative (HND) S-H bonds, first row in the table, are always energetically more stable than all dissociative structures by at least 0.2 eV per BDT throughout all systems investigated. Although an isolated Au adatom is frequently seen on Au(111), other contacts, e.g. with perfect surface or small Au cluster, do emerge in a molecular junction. Both our calculation and a previous DFT total energy calculation [29] prefer the HND configuration, rather than the HD configuration, in the presence of either small cluster or perfect (111) surface, which is consistent with the experimental observation [28]. Furthermore, as elucidated below, our calculated breakdown force of the HND junction (by elongation) is 1.38 nN, fairly close to the experimental value (1.6 ± 0.2 nN) of similar molecular junction with the same thiol linker [18]. These results strongly suggest that an HND structure is worth of investigation, and may provide a more realistic model in terms of transport modeling of BDT devices.

### 3.2 Evolution of junctions

Geometry of the scattering region of all instances of Au/BDT/Au junctions, in a ``dual-slab'' model, is again fully optimized with VASP. An adatom is available on each of the two face-to-face surfaces of a junction, while BDT molecule, in HND (Fig. 3(a)) or HD (Fig. 3(b)) case, was put in between the two adatoms. When the H is dissociated, the Au-S bond shows a stronger strength than the Au-Au bond in our calculation. Therefore, in STM experiments where a STM tip was repeatedly retracted from an Au surface, it is quite likely to drag an Au atomic wire out of the surface through the attached BDT molecule [2]. To cover this possibility, for the HD structure, we have calculated a few devices where the H-dissociated BDT is linked to the surface via gold atomic chains with various lengths (Fig. 3(c)). The parallel configurations, as shown in Fig. 3(a), were adopted to build the dual-slab model of the junctions, since they have lower energies than the perpendicular ones.

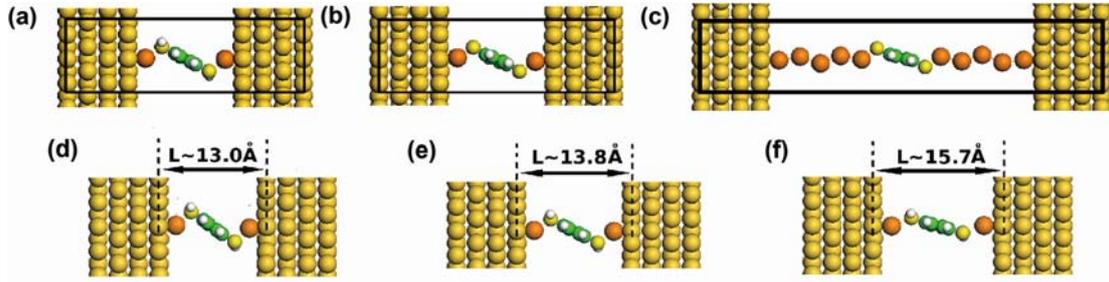

Fig. 3 Atomic models of the (a) non-dissociated and (b) dissociated BDT linked by ad-atoms and (c) gold atomic chains. Fully relaxed configurations for H-non-dissociative junctions, (d) at $L = 13.0$ Å, in which $L_L = 6.846$ Å $\approx L_R = 6.849$ Å, very close to the $D$ value of an adatom on a bare surface; (e) the junction is at its equilibrium, when $L = 13.8$ Å, in which $L_L = 6.935$ Å $\approx L_R = 6.933$ Å; (f) the junction starts to break at $L = 15.7$ Å, $L_L = 7.101$ Å, $L_R = 7.052$ Å.

The H-non-dissociated junctions always have lower total energies than the corresponding H-dissociated cases: because there are two Au/BDT interfaces now, the energy difference roughly doubles those listed in Table I. The total energy of the dual-slab structure as a function of the width of junction gap $L$ are shown in Fig. 4(a). In the HND case, the energy goes lower and lower as one slab is moved further from $L = 13.0$ Å ( Fig. 3(d)) and reaches a minimum value at around $L = 13.8$ Å (Fig. 3(e)). This value of $L$ is regarded as the equilibrium width of the junction, in which the Au-S bond length is 2.43 Å and the $L_L$ ($L_R$) is 6.94 Å. Near $L = 15.7$ Å (structure shown in Fig. 3(f)), the junction starts to break which can be identified from either geometric or energetic points of view. In terms of energetics, the total energy keeps essentially constant from 15.7 Å to 15.8 Å indicating the junction breaking. Furthermore, distances $L_L$ and $L_R$ are fairly identical before $L$ reaches 15.6 Å. However, an appreciable difference of 0.05 Å between them becomes observable at $L = 15.7$ Å, and it becomes 0.08 Å for $L = 15.8$ Å. Both the Au-S bond length (left lead) and the $L_R$ value start to decrease at $L=15.7$ Å. All these geometrical changes are consistent with the features of a junction breaking down, in agreement with the energetic behavior. Structural details of this series of junctions are available in Table II. It is remarkable that the values of $L_L$ (6.846 Å) and $L_R$ (6.849 Å) in the configuration for $L$ = 13.0 Å are quite close to the $D$ value of 6.846 for an adatom on a bare Au(111) surface, which implies that the absolute tension applied to the adatom in the $z$ direction should be the smallest (close to zero) when the junction length is around 13.0 Å.

Table II. Changes of $L_L$, $L_R$, and Au-S bond lengths at the left (Au-S(L)) and right (Au-S(R)) leads with respect to (H-non-dissociative) junction length $L$. All values are in Angstrom

| $L$ | 12.7 | 12.8 | 12.9 | 13.0 | 13.1 | 13.2 | 13.5 | 13.6 | 13.7 | 13.8 | 13.9 |
|---|---|---|---|---|---|---|---|---|---|---|---|
| $L_L$ | 6.823 | 6.834 | 6.841 | 6.849 | 6.854 | 6.866 | 6.903 | 6.915 | 6.930 | 6.935 | 6.946 |
| $L_R$ | 6.817 | 6.833 | 6.838 | 6.846 | 6.857 | 6.874 | 6.900 | 6.910 | 6.917 | 6.933 | 6.942 |
| Au-S (L) | 2.426 | 2.426 | 2.422 | 2.420 | 2.418 | 2.419 | 2.422 | 2.421 | 2.423 | 2.425 | 2.429 |
| Au-S (R) | 2.430 | 2.429 | 2.425 | 2.422 | 2.421 | 2.422 | 2.422 | 2.421 | 2.423 | 2.427 | 2.429 |

| $L$ | 14.0 | 14.1 | 14.2 | 14.7 | 15.2 | 15.3 | 15.4 | 15.5 | 15.6 | 15.7 | 15.8 |
|---|---|---|---|---|---|---|---|---|---|---|---|
| $L_L$ | 6.958 | 6.968 | 6.979 | 7.030 | 7.078 | 7.087 | 7.094 | 7.095 | 7.101 | 7.101 | 7.103 |
| $L_R$ | 6.953 | 6.965 | 6.977 | 7.035 | 7.078 | 7.093 | 7.098 | 7.100 | 7.093 | 7.052 | 7.023 |
| Au-S (L) | 2.431 | 2.434 | 2.439 | 2.474 | 2.535 | 2.543 | 2.562 | 2.567 | 2.568 | 2.539 | 2.534 |
| Au-S (R) | 2.434 | 2.435 | 2.442 | 2.476 | 2.534 | 2.572 | 2.592 | 2.638 | 2.704 | 2.886 | 3.017 |

## 3.3 Breakdown force

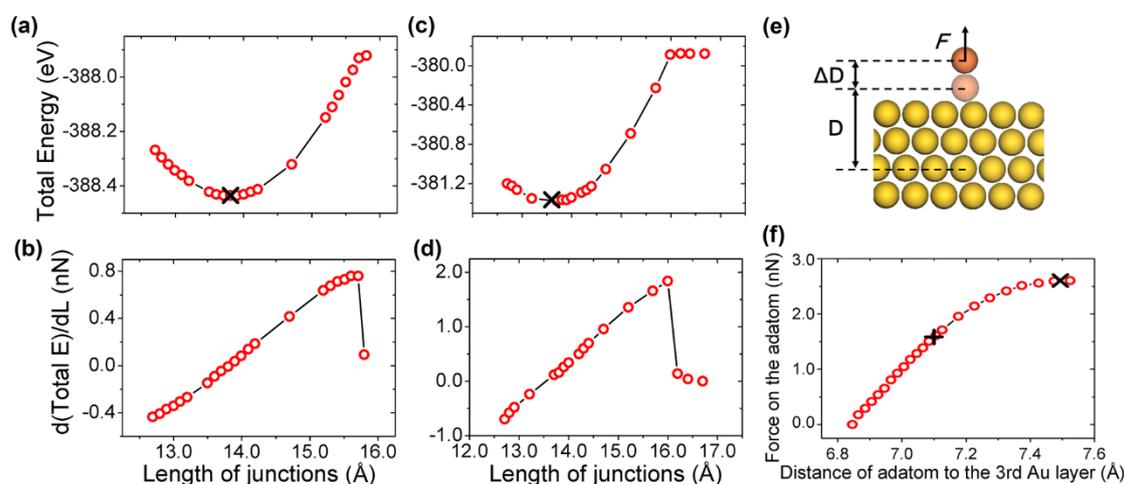

Fig. 4. Calculated (red dots) and fitted (black lines) data for H-non-dissociative (a), (b) and H-dissociative (c), (d) junctions of: (a), (c) total energies of the junction with different lengths, the cross indicates the equilibrium position; and (b), (d) derived forces from curve (a) and (c) according to equation $F_{Junction} = dE/dL \approx \Delta E/\Delta L$. The second force calculation method is illustrated in (e) where a force is acted on the adatom at non-equilibrium positions in the transport direction. (f) shows the results of this force calculation in which the largest likely $D$ value obtained from the H-non-dissociative junction elongation is marked by a ``plus'' symbol; that from the H-dissociative junction elongation is marked by a ``cross''.

In our calculations, two methods were employed to estimate the changes of tension with respect to the junction elongation, i.e. (1) derive the force from total energy of the elongating junction and, (2) directly calculate the force according to the shift of the Au adatom. In the first method, the tension of the junction is derived by $F_{Junction} = dE/dL \approx \Delta E/\Delta L$. Figure 4(b) plots the calculated $F_{Junction}$ versus $L$, where its value at $L=13.8$ Å (the equilibrium junction length) is zero. The largest positive value of this force is 0.76 nN, but the lower limit of negative forces is rather ambiguous. At $L = 13.0$ Å, the values of $L_L$ and $L_R$ are very close to that of $D$ (of a bare surface) at equilibrium. It is therefore quite reasonable to use that junction length as a reference in estimating the lowest negative force. The force of the junction with $L = 13.0$ Å is -0.34 nN as shown in Fig. 4(b), we thus obtain an estimated breakdown force of 1.10 nN.

The second method is to estimate the force according to the vertical position of an adatom on a bare Au surface as shown in Fig. 4(e). After the breakdown of a two-probe junction is completed, the adatom on the surface of a lead, without molecules attached, is retracted from its longest stretching position to its equilibrium position on the surface. At the equilibrium position, the tension is zero by definition; while at the longest stretching position just before breakdown starts, i.e. at $L = 15.6$ Å from Table II, the force can be directly evaluated. The breakdown force is thus equal to the acting force directly calculated from the adatom on a bare surface in the $z$ direction, in which the $D$ is increased to match $L_L$ or $L_R$ of the junction at $L = 15.6$ Å. Table II shows that $L_L \approx L_R \approx 7.10$ Å at $L = 15.6$ Å, giving rise to a value of the breakdown force of 1.58 nN with $D = 7.10$ Å, as shown by the black ``plus'' in Fig. 4(f).

The two methods for calculating breakdown force introduce some small uncertainty either to the minimum or to the maximum tension due to the HS-Au bonding (e.g. charge redistribution), but in different ways. For the first method, the calculated tension before the junction breakdown is exact (within the DFT technique); but this tension is underestimated after the junction breakdown because the first method included the HS-Au bonding that tends to somewhat lower the breakdown force. For the second method the situation is opposite, namely it overestimates the breakdown force without considering the influence of HS-Au bonding at the longest stretching position before the breakdown. These considerations imply that the correct theoretical value should be in between the values obtained by our two methods, i.e. in the range between 1.10 nN to 1.58 nN. This result is consistent with the experimental value, i.e. 1.6 ± 0.2 nN, of another molecule connected to Au leads with the same thiol linker [18].

As a comparison, we also evaluated the breakdown force of the H-dissociative model following the same procedure. Figure 4(c) shows the evolution of total energy as a function of $L$. Its derivative curve, representing the tension of the junction as a function of $L$, is shown in Fig. 4(f). The zero value of the tension locates at L=13.6 Å, while the largest positive value is clearly shown as 1.85 nN. The lowest negative force was determined using the same idea that at $L = 12.8$ Å, the force is found -0.59 nN as shown in Fig. 4(d). Therefore, a breakdown force of 1.85 nN + 0.59 nN = 2.44 nN is obtained for the H-dissociative junction. The adatom-to-surface distance at the longest stretching position (at $L = 16.0$ Å), just before breakdown starts, must be ascertained so that the breakdown force can be estimated using the second method. It was found that, according to the fully relaxed atomistic structure, $L_L \approx L_R \approx 7.495$ Å at $L = 16.0$ Å. The force is 2.60 nN when $D = 7.495$ Å, as shown in Fig. 4(f) indicated by a black cross. The calculated breakdown force for the H-dissociative junctions, in the range from 2.44 nN to 2.60 nN, is significantly higher than the experimentally measured value of the thiol-gold linker

[18] which is 1.6 ± 0.2 nN. On the other hand, the theoretical breakdown force for H-non-dissociative junctions are found in between 1.10 nN and 1.58 nN which is more consistent to the measured data. These results suggest that the H-non-dissociative model gives a more reasonable description for the interfacial contact in Au/BDT/Au experiments [1-3].

**3.4 Transport Properties**

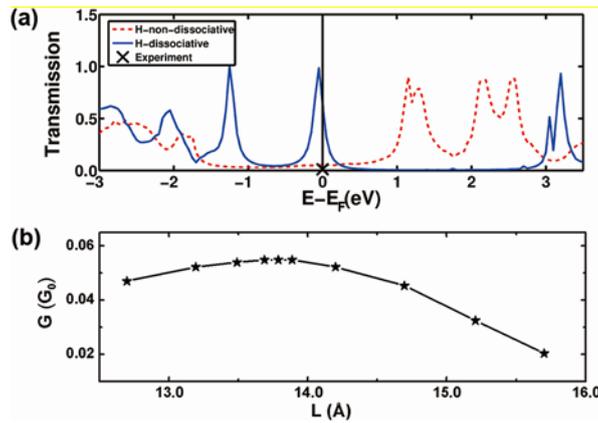

Fig. 5 (a) Transmission of H-dissociative (solid blue) and H-non-dissociative (dashed red) models versus energy without junction stretching. The cross denotes the experimental value 0.011 $G_0$ [2]. (b) Conductance ($G_0$) versus junction length (Å) under mechanical stretching for H-non-dissociative model.

Figure 5(a) shows the transmission (*T*) spectra of Au/BDT/Au devices versus energy *E*, in a range of -3.0 eV to 3.5 eV. For the HND model, the equilibrium conductance (the value of *T* at the Fermi level) is 0.054 $G_0$ for a junction at its equilibrium junction length *L* = 13.8 Å. This conductance decreases when stretching the junction, and reaches 0.02 $G_0$ for *L* ~ 15.7 Å, at which the junction starts its mechanical breakdown. Figure 5(b) plots conductance changes versus *L*. This range of conductance, 0.02-0.054 $G_0$, is within a factor of two to five in comparison to the experimental value of 0.011 $G_0$ [2] to 0.0132 $G_0$ [1]. On the other hand, the conductance of all HD models with or without stretching are much higher, in a range of 0.38-0.86 $G_0$. These high values are consistent with previous *ab initio* calculations [12-15] which used HD models. Inclusion of many-body effects, e.g. calculations done with many-body perturbation theory, reduces the equilibrium conductance of the NHD model to 0.02-0.03 $G_0$, by a factor of 2 at the equilibrium junction length, however, enlarges that of the HD model to an even larger value [30,31].

### 3.5 Bonding Mechanism

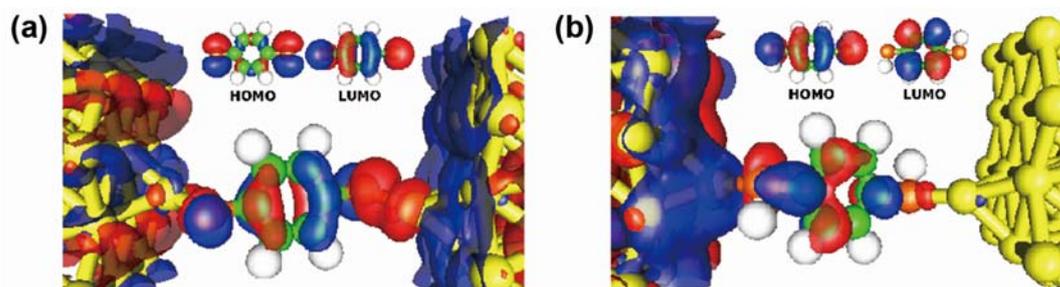

Fig. 6 Scattering states of (a) H-dissociative and (b) H-non-dissociative models. Insets are the HOMO and LUMO of the corresponding molecule 1,4-benzenedithiolate and 1,4-benzenedithiol.

The transport results clearly suggest that the hybridization of electronic states from Au electrodes and the molecule is significantly different for H-non-dissociative and dissociative models, and thus leads to diverse behaviors of the conductance evolution with respect to the length of junction. To find the reason why this happens, we have analyzed the scattering states around the Fermi level for both models, as shown in Fig. 6. When the H is dissociated from an S-H group, an electron of the S atom becomes unpaired which has an overwhelming tendency to attract an additional electron to make a pair. The additional electron is most likely contributed by s-electrons of Au leads, resulting in a transfer of charge from leads to the molecule. The transferred electron dopes into the lowest unoccupied molecular orbital (LUMO) of 1,4-benzenedithiolate and pushes down the *s*-LUMO bonding state just below the Fermi level, as found in a similar system [32,33]. The *s*-LUMO bonding states are expected to be much delocalized, since it was composed of a delocalized LUMO and a delocalized metal *s* state. We thus plot the scattering states around the Fermi level, as shown in Fig. 6(a). It was found that the conductance around the Fermi level is indeed dominated by a delocalized *s*-LUMO state as expected. The plot shows the scattering states nicely passing through the junction, giving rise to a high conductance value. On the other hand, for the H-non-dissociative model, the above charge transfer can hardly happen since all electrons are paired already, hence the hybridized *s*-LUMO state around the Fermi level disappears. The bonding picture then switches to that of a lone-pair of the S atom donating to the partially unfilled *s*-band of Au leads. The lone-pair is rather localized, like a σ-type orbital. Similar dative bonding mechanism was also found in thiophene/Cu interfaces [34]. The hybridization between the lone-pair and Au leads is therefore somewhat localized, which results in a tunneling mechanism for electrons going through the junction at low bias; a much smaller conductance is therefore expected. Indeed, as shown in the plot of scattering states in Fig. 6(b), very few incoming scattering states can pass through the junction. In this case, the conductance is mainly contributed by the HOMO of 1,4-benzenedithiol. These results unambiguously manifest that the interfacial bonding configuration, rather than if considering the many-body effects, is curial to properly predict the equilibrium conductance of molecular junctions. With the established correlation of bonding mechanism and equilibrium conductance, it can be inferred that the conductance should be similar, but slightly smaller, if replacing the BDT molecule by 1,4-benzenediamine (BDA). It is because that the $NH_2$ group bonds to the Au lead through an electron lone pair of N and it is more localized owing to higher electron affinity of N. This hypothesis is well supported by experiments with BDA and other linker groups [35].

## 4 CONCLUSIONS

In summary, we showed there are two bonding mechanism for the Au-S bond in an Au/BDT/Au junction, namely a covalent bond filled by electrons transferred from Au lead to S atom and a dative bond formed by a lone pair from the S atom. These two bonding configurations were experimentally confirmed by scanning tunneling microscopy with atomic manipulation and were manifested to have different energy level alignments [36] as suggested in this work. We have demonstrated that the mechanical and electric properties of a molecular junction have strong correlation with the bonding configuration at the contact interface. Another bonding mechanism was revealed in this work, surprisingly, that the H atoms in the thiol group of the BDT energetically prefers to be non-dissociative after the Au/BDT/Au transport junctions are formed. The introduction of non-dissociative H atoms, in other words the dative bonding configuration, blocks charge transfer doping to the BDT from the Au electrodes, effectively induces an extra potential barrier that considerably reduces the electron transparency of the Au/BDT interface. It had been used to believe that the HD model, namely the delocalized covalent bonding picture, is the right model for an Au/BDT/Au junction. With substantial efforts from our work and the promoted successive studies, however, the conductance was predict, with the HND model, within a factor of two to five in comparison with experimental values [1,2]. The factor of two to five reduces to less than a factor of two if the many-body effect is considered, while the inclusion of many-body effects leads to an even larger conductance for the HD model [30,31].

We speculate that these two bonding configurations are switchable between each other by changing the local chemical environment, e.g. local geometry [37], at the interface, so that the electric conductance is thus, most likely, tunable by the interfacial environment. Our investigation reveals how the interfacial geometry correlates with the interfacial bonding configuration and could drastically influence transport properties. These findings shed considerable light on charge conduction properties at the single molecule level.

**Acknowledgements**

We thank Dr. Lei Liu for his assistance concerning the NEGF-DFT code and Prof. M. A. Reed for helpful discussions on experimental issues of the junction. This work was financially supported by NSERC of Canada, FQRNT of Quebec and CIFAR (H.G.) for work done in Canada; and the National Natural Science Foundation of China (NSFC) under Grant No. 11004244 and 11274380, the National Basic Research Program of China under grant No. 2012CB932704, the Beijing Natural Science Foundation (BNSF) under Grant No. 2112019, and the Basic Research Funds in Renmin University of China from the Central Government under Grant No. 12XNLJ03 (W.J.) for work done in China. We are grateful to RQCHP for providing computational facilities.

1. H. Song, Y. Kim, Y. H. Jang, H. Jeong, M. A. Reed and T. Lee, Observation of molecular orbital gating, *Nature*, 2009, 462(7276): 1039
2. Xiao, Xu and N. J. Tao, Measurement of Single Molecule Conductance: Benzenedithiol and Benzenedimethanethiol, *Nano Lett.*, 2004, 4(2): 267
3. M. Tsutsui, M. Taniguchi and T. Kawai, Atomistic Mechanics and Formation Mechanism of Metal–Molecule–Metal Junctions, *Nano Lett.*, 2009, 9(6): 2433
4. M. Di Ventra, S. T. Pantelides and N. D. Lang, The benzene molecule as a molecular resonant-tunneling transistor, *Applied Physics Letters*, 2000, 76(23): 3448
5. K. Stokbro, J. Taylor, M. Brandbyge, J. L. Mozos and P. Ordejón, Theoretical study of the nonlinear conductance of Di-thiol benzene coupled to Au(1 1 1) surfaces via thiol and thiolate bonds, *Computational Materials Science*, 2003, 27(1–2): 151
6. T. Tada, M. Kondo and K. Yoshizawa, Green's function formalism coupled with Gaussian broadening of discrete states for quantum transport: Application to atomic and molecular

wires, *J. Chem. Phys.*, 2004, 121(16): 8050

7. S.-H. Ke, H. U. Baranger and W. Yang, Molecular Conductance: Chemical Trends of Anchoring Groups, *Journal of the American Chemical Society*, 2004, 126(48): 15897
8. P. Delaney and J. C. Greer, Correlated Electron Transport in Molecular Electronics, *Phys. Rev. Lett.*, 2004, 93(3): 036805
9. G. C. Solomon, J. R. Reimers and N. S. Hush, Overcoming computational uncertainties to reveal chemical sensitivity in single molecule conduction calculations, *J. Chem. Phys.*, 2005, 122(22):
10. R. B. Pontes, F. D. Novaes, A. Fazzio and A. J. R. da Silva, Adsorption of Benzene-1,4-dithiol on the Au(111) Surface and Its Possible Role in Molecular Conductance, *Journal of the American Chemical Society*, 2006, 128(28): 8996
11. D. Q. Andrews, R. P. Van Duyne and M. A. Ratner, Stochastic Modulation in Molecular Electronic Transport Junctions: Molecular Dynamics Coupled with Charge Transport Calculations, *Nano Lett.*, 2008, 8(4): 1120
12. J. Nara, W. T. Geng, H. Kino, N. Kobayashi and T. Ohno, Theoretical investigation on electron transport through an organic molecule: Effect of the contact structure, *J. Chem. Phys.*, 2004, 121(13): 6485
13. C. Toher and S. Sanvito, Efficient Atomic Self-Interaction Correction Scheme for Nonequilibrium Quantum Transport, *Phys. Rev. Lett.*, 2007, 99(5): 056801
14. C. Toher and S. Sanvito, Effects of self-interaction corrections on the transport properties of phenyl-based molecular junctions, *Phys. Rev. B*, 2008, 77(15): 155402
15. M. Strange, I. S. Kristensen, K. S. Thygesen and K. W. Jacobsen, Benchmark density functional theory calculations for nanoscale conductance, *J. Chem. Phys.*, 2008, 128(11):
16. S. Y. Quek, H. J. Choi, S. G. Louie and J. B. Neaton, Length Dependence of Conductance in Aromatic Single-Molecule Junctions, *Nano Lett.*, 2009, 9(11): 3949
17. M. A. Reed, C. Zhou, C. J. Muller, T. P. Burgin and J. M. Tour, Conductance of a Molecular Junction, *Science*, 1997, 278(5336): 252
18. Huang, Xu, Chen, M. D. Ventra and Tao, Measurement of Current-Induced Local Heating in a Single Molecule Junction, *Nano Lett.*, 2006, 6(6): 1240
19. J. P. Perdew, K. Burke and M. Ernzerhof, Generalized Gradient Approximation Made Simple, *Phys. Rev. Lett.*, 1996, 77(18): 3865
20. G. Kresse and D. Joubert, From ultrasoft pseudopotentials to the projector augmented-wave method, *Phys. Rev. B*, 1999, 59(3): 1758
21. G. Kresse and J. Furthmüller, Efficient iterative schemes for *ab initio* total-energy calculations using a plane-wave basis set, *Phys. Rev. B*, 1996, 54(16): 11169
22. J. Taylor, H. Guo and J. Wang, *Ab initio* modeling of quantum transport properties of molecular electronic devices, *Phys. Rev. B*, 2001, 63(24): 245407
23. Z. Ning, Y. Zhu, J. Wang and H. Guo, Quantitative Analysis of Nonequilibrium Spin Injection into Molecular Tunnel Junctions, *Phys. Rev. Lett.*, 2008, 100(5): 056803
24. Y. Hu, Y. Zhu, H. Gao and H. Guo, Conductance of an Ensemble of Molecular Wires: A Statistical Analysis, *Phys. Rev. Lett.*, 2005, 95(15): 156803
25. M. Kamenetska, M. Koentopp, A. C. Whalley, Y. S. Park, M. L. Steigerwald, C. Nuckolls, M. S. Hybertsen and L. Venkataraman, Formation and Evolution of Single-Molecule Junctions, *Phys. Rev. Lett.*, 2009, 102(12): 126803
26. C.-C. Kaun and H. Guo, Resistance of Alkanethiol Molecular Wires, *Nano Lett.*, 2003, 3(11):

1521

27. F.-S. Li, W. Zhou and Q. Guo, Uncovering the hidden gold atoms in a self-assembled monolayer of alkanethiol molecules on Au(111), *Phys. Rev. B*, 2009, 79(11): 113412
28. I. I. Rzeźnicka, J. Lee, P. Maksymovych and J. T. Yates, Nondissociative Chemisorption of Short Chain Alkanethiols on Au(111), *J. Phys. Chem. B*, 2005, 109(33): 15992
29. J.-G. Zhou and F. Hagelberg, Do Methanethiol Adsorbates on the Au(111) Surface Dissociate?, *Phys. Rev. Lett.*, 2006, 97(4): 045505
30. T. Rangel, A. Ferretti, P. E. Trevisanutto, V. Olevano and G. M. Rignanese, Transport properties of molecular junctions from many-body perturbation theory, *Phys. Rev. B*, 2011, 84(4): 045426
31. M. Strange, C. Rostgaard, H. Häkkinen and K. S. Thygesen, Self-consistent GW calculations of electronic transport in thiol- and amine-linked molecular junctions, *Phys. Rev. B*, 2011, 83(11): 115108
32. W. Ji, Z.-Y. Lu and H.-J. Gao, Multichannel interaction mechanism in a molecule-metal interface, *Phys. Rev. B*, 2008, 77(11): 113406
33. W. Ji, Z.-Y. Lu and H. Gao, Electron Core-Hole Interaction and Its Induced Ionic Structural Relaxation in Molecular Systems under X-Ray Irradiation, *Phys. Rev. Lett.*, 2006, 97(24): 246101
34. Z.-X. Hu, H. Lan and W. Ji, Role of the dispersion force in modeling the interfacial properties of molecule-metal interfaces: adsorption of thiophene on copper surfaces, *Sci. Rep.*, 2014, 4:
35. L. Venkataraman, J. E. Klare, C. Nuckolls, M. S. Hybertsen and M. L. Steigerwald, Dependence of single-molecule junction conductance on molecular conformation, *Nature*, 2006, 442(7105): 904
36. Y. Jiang, Q. Huan, L. Fabris, G. C. Bazan and W. Ho, Submolecular control, spectroscopy and imaging of bond-selective chemistry in single functionalized molecules, *Nat. Chem.*, 2013, 5(1): 36
37. F. Cheng, W. Ji, L. Leung, Z. Ning, J. C. Polanyi and C.-G. Wang, How Adsorbate Alignment Leads to Selective Reaction, *ACS Nano*, 2014, 8(8): 8669